%% file: main.tex
\documentclass[twoside]{article}

\usepackage[accepted]{aistats2020}
\usepackage[round]{natbib}

\usepackage[utf8]{inputenc}
\usepackage[T1]{fontenc}
\usepackage{hyperref}
\usepackage{url}
\usepackage{amsfonts,amssymb,amsthm,amsmath}
\usepackage{nicefrac}
\usepackage{microtype}
\usepackage[dvipsnames]{xcolor}
\usepackage{graphicx}
\usepackage{bm}
\usepackage[inline]{enumitem}
\usepackage[noend]{algorithm2e}
\usepackage{tikz}
\usetikzlibrary{decorations.pathmorphing,decorations.shapes}
\usepackage{etoolbox} 
\usepackage{mathtools}
\usepackage[disable]{todonotes}
\usepackage{cleveref}  

\input{macros}

\graphicspath{{./fig/}}

\setitemize{leftmargin=*}
\setlist{nosep}

\begin{document}

%

%
\runningauthor{Hron, Krauth, Jordan, Kilbertus}

\twocolumn[

\aistatstitle{Exploration in two-stage recommender systems}

\aistatsauthor{Jiri Hron${}^{1,*}$, Karl Krauth${}^{2,*}$, Michael I.~Jordan${}^2$, Niki Kilbertus${}^{1,3}$}

\aistatsaddress{${}^1$University of Cambridge, ${}^2$UC Berkeley, ${}^3$Max Planck Institute for Intelligent Systems
}
]

\begin{abstract}
\input{content/00abstract}
\end{abstract}

\input{content/01intro}

\input{content/02setup}
\input{content/03algo}


\input{content/05conclusion}


\bibliography{ref}
\bibliographystyle{refstyle}


\end{document}

%% file: macros.tex
\DeclareMathOperator*{\argmax}{\arg\!\max}
\DeclarePairedDelimiterX{\infdiv}[2]{(}{)}{%
  #1\;\delimsize\|\;#2%
}

\newcommand{\E}{\mathbb{E}}
\newcommand{\R}{\mathbb{R}}
\newcommand{\N}{\mathbb{N}}
\newcommand{\gauss}{\mathcal{N}}
\newcommand{\given}{\, | \,}

\definecolor{SteelBlue}{RGB}{70, 130, 180}

\newcommand{\hrulespace}[1]{
\vspace{#1}
\hrule
\vspace{#1}
}
\newtoggle{showtodos}
\toggletrue{showtodos}

\iftoggle{showtodos}{
    \newcommand{\nk}[1]{\todo[color=red!40]{NK: #1}}
    \newcommand{\kk}[1]{\todo[color=blue!40]{KK: #1}}
    \newcommand{\jh}[1]{\todo[color=cyan!40]{JH: #1}}
    \hypersetup{draft}
}{
    \newcommand{\nk}[1]{}
    \newcommand{\kk}[1]{}
    \newcommand{\jh}[1]{}
}

%% file: content/00abstract.tex
\noindent%
Two-stage recommender systems are widely adopted in industry due to their scalability and maintainability.
These systems produce recommendations in two steps:
(i)~multiple \emph{nominators} preselect a small number of items from a large pool ($\gg$100K) using cheap-to-compute item embeddings; 
(ii)~with a richer set of features, a \emph{ranker} rearranges the nominated items and serves them to the user.
A key challenge of this setup is that optimal performance of each stage in isolation does not imply optimal global performance.
In response to this issue, \citet{ma2020off} proposed a nominator \emph{training objective} importance weighted by the ranker's probability of recommending each item.
In this work, we focus on the complementary issue of \emph{exploration}.
Modeled as a contextual bandit problem, we find LinUCB---a near optimal exploration strategy for single-stage systems---may lead to linear regret when deployed in two-stage recommenders.
We therefore propose a method of \emph{synchronising the exploration strategies} between the ranker and the nominators.
Our algorithm only relies on quantities already computed by standard LinUCB at each stage and can be implemented in three lines of additional code.
We end by demonstrating the effectiveness of our algorithm experimentally.

%% file: content/01intro.tex

\section{Introduction}
\label{sec:intro}

Contemporary recommender systems are tasked with finding a small number of relevant items among millions or billions of candidates, personalized for each of hundreds of thousands or millions of users and their always changing needs, all of which has to happen in order of milliseconds so as not to negatively impact webpage loading speeds.
One of the most widely used solutions to the problem are \emph{two-stage recommender systems} \citep{borisyuk2016casmos,covington2016deep,eksombatchai2018pixie} in which (i)~a set of \emph{computationally efficient} nominators narrows down the search from millions to only hundreds of items, and (ii)~the \emph{slower but more accurate} ranker selects and reorders a few items which are eventually served to the user.

For example, a nominator can use a two-tower architecture \citep{yi2019sampling} and focus on a narrower set of features, whereas the ranker would rely on a more powerful model and consider additional features extracted from, e.g., ratings, specialized user and item attributes, or the number and type of past interactions with given user \citep{covington2016deep,ma2020off}.
Importantly, the nominators are often heterogeneous both in terms of the size and type of the items from which they select the candidate items, and the algorithm used to select candidates ranging from simple associative rules to recurrent neural networks \citep{chen2019top}.\jh{I'm pretty sure I'm missing citations throughout; please do not hold back from adding as many as you can.}
We will focus on nominators which utilize statistical learning methods in the two-stage setup as in the paper most relevant to our work \citep{ma2020off}.

\citet{ma2020off} study off-policy learning for two-stage recommender systems where the goal is learning a good recommendation policy from the typically abundant logged data. 
The main proposal of \citeauthor{ma2020off}\ is to modify the nominator training objective by adding importance weights based on the ranker's probability of recommending each item.
With adjustments facilitating gradient descent optimization, the authors show significant empirical improvements not only compared to a system trained without importance weighting, but also relative to nominators importance weighted only based on the past \emph{nominator} policy (ignoring the presence of the ranker).
These results thus demonstrate that local optima of individual components do not translate to optimality of the system as a whole.

Naturally, we can ask whether there are other aspects of the recommendation problem where optimal solutions for a single-stage system result in suboptimal performance when deployed in a two-stage system.
We answer this question affirmatively in the case of \emph{exploration} by which we mean the task of learning an optimal recommendation policy under uncertainty in a sample efficient way.
An effective exploration strategy then needs to balance greedy actions based on past interactions with exploratory recommendations targetting items about which there is little or no information.

While many strategies have been proposed in the literature, ranging from more conservative ones like Boltzmann exploration \citep{daw2006cortical,chen2019top} to more optimistic contextual bandit algorithms \citep{lattimore2020bandit}, we will restrict our discussion to the popular LinUCB algorithm \citep{auer2002using,dani2008stochastic,li2010contextual} and the associated contextual bandit recommendation setup. 
A similar setup has been explored by \citep{ma2020off} as it is rich enough to exhibit many of the salient properties encountered in the real world, while abstracting away some of the complexities involved in deployment of large scale recommendation systems.

\textbf{Contributions:} We
(i)~show that a mismatch of feature mappings (architecture) and the amount of data seen (regularization) between LinUCB ranker and nominators can result in large, even linear regret;
(ii)~demonstrate said effects and their dependence on the level of mismatch empirically on a toy dataset; and
(iii)~propose a simple algorithm based on synchronization of inferred statistics between the ranker and the nominators which addresses the issue. We demonstrate the efficacy of our algorithm on simulated data.



%% file: content/02setup.tex
\section{Setup}
\label{sec:setup}

\begin{figure}
  \centering
  \includegraphics[width=0.7\columnwidth]{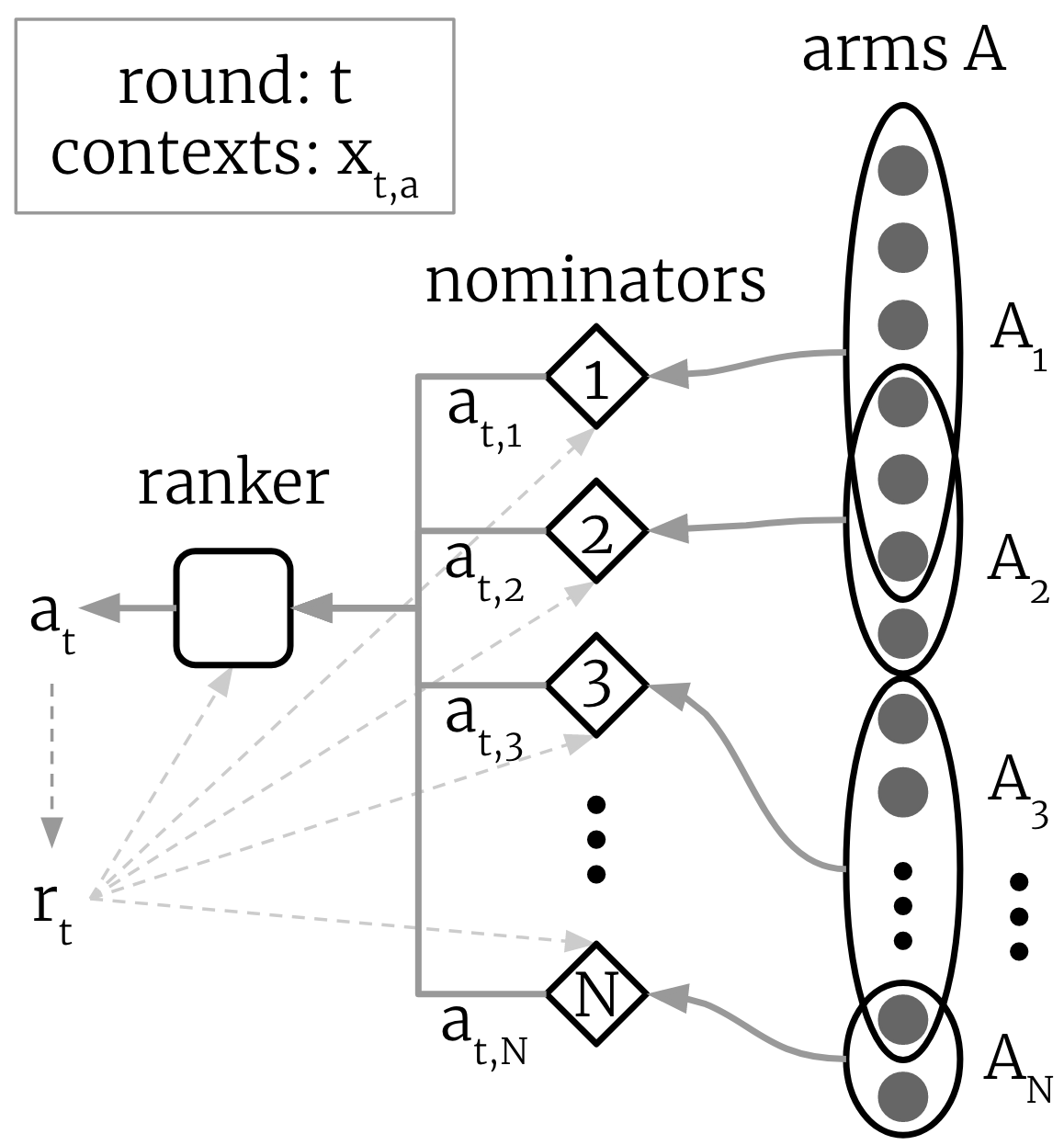}
  \caption{The two-stage recommendation setup.}
  \label{fig:setup}
\end{figure}

We consider a scenario in which a single item is to be recommended in each of the $T$ rounds.
The recommendation problem is modeled in a contextual bandit setup where the individual items correspond to arms $a \in A$ (we thus also refer to items as arms or actions).
For the first stage, we assume a fixed number of $N \in \N$ \emph{nominators}, each of which has access only to a fixed non-empty subset of arms $A_n \subseteq A$.
At every round $t \in [T]$, each nominator observes contexts $x_{t, a}$ for all $a \in A_n$, and selects a \emph{single} action $a_{t, n}$.
The ranker then chooses a \emph{single} final recommendation among the ones nominated in the first stage $a_t \in \{ a_{t, 1}, \ldots a_{t, N} \}$ \emph{based on the corresponding contexts}.\jh{Maybe cite Singla et al.\ (LEARNING WITH LEARNING EXPERTS) and Aggarwall et al. (CORALL) and explain that this is the main distinction of our setup from theirs?!}
Both the ranker and the nominators can ultimately be updated using the reward $r_t$ obtained by pulling arm $a_t$ as well as all the revealed contexts $x_{t, a}$.
This two-stage recommendation process is illustrated in \Cref{fig:setup}.

We restrict our attention to the stochastic linear bandit setting \citep{abe1999associative,lattimore2020bandit}, and the LinUCB algorithm \emph{with ellipsoidal confidence sets} for both the ranker and the nominators \citep{auer2002using,dani2008stochastic}.
The linear contextual bandit setting assumes existence of a fixed embedding for each context $\phi(x_{t , a})$ such that
\begin{equation}
  \E [ r_t \given x_{t, a_t} , a_t ] = \langle \phi(x_{t, a_t}) , \theta_\star \rangle
\end{equation}
for all $x_{t, a}$, $a_t$, and a fixed $\theta_\star \in \R^{d}$.
Since $\theta_\star$ is assumed unknown, a \emph{single-stage} LinUCB estimates it by ridge regression with regularization parameter $\lambda > 0$
\begin{align}
\begin{split}\label{eq:blr_posterior_params}
    \Sigma_t 
    &\coloneqq
    \Bigl[ \lambda\, I_{d} + \sum_{i = 1}^t \phi(x_{i, a_i})  \phi(x_{i, a_i})^\top\Bigr]^{-1} 
    \, ,
    \\
    \hat{\theta}_t 
    &\coloneqq 
    \Sigma_t \sum_{i = 1}^t r_i\, \phi(x_{i, a_i})
    \, .
\end{split}
\end{align}
The actions are then selected according to
\begin{align}\label{eq:action_selection}
    a_{t+1} &\in \argmax_{a \in A}\, \text{UCB}_{t+1} (a)
    \, ,
\end{align}
where as in \citep[p.~239--241]{lattimore2020bandit}
\begin{align*}
    \text{UCB}_{t+1} (a)
    &\coloneqq
    \langle \phi(x_{t, a}), \hat{\theta}_t \rangle + \sqrt{\beta_t}\, \| \phi(x_{t, a}) \|_{\Sigma_t}
    \, ,
    \\
    \sqrt{\beta_t}
    &\coloneqq
    \sqrt{\lambda}
    +
    \sqrt{
      2 \log t
      +
      d
      \log \left(
        \frac{d \lambda + t}{d \lambda}
      \right)
    }
    \, .
\end{align*}
LinUCB with such $\beta_t$ achieves near optimal regret
\begin{align}\label{eq:regret_definition}
\begin{aligned}[c]
    R_T 
    &\coloneqq
    \E \left[
      \sum_{t=1}^T
        r_{t,\star} - r_t
    \right]
    \\
    &=
  \sum_{t=1}^T
    \langle 
      \theta_\star ,
      \phi(x_{t, a_{t, \star}})
      -  
      \E [\phi(x_{t, a_t})]
    \rangle
  \, ,
\end{aligned}
\end{align}
when the reward noise is sub-Gaussian \citep{dani2008stochastic,lattimore2020bandit},
where
\begin{equation}
  a_{t, \star} \coloneqq \argmax_{a \in A} \langle \theta_\star , \phi (x_{t, a}) \rangle \, ,
\end{equation}
$r_{t, \star}$ is the reward obtained by choosing $a_{t, \star}$, and the expectation is taken with respect to the randomness of the rewards and the policy (uniform tie breaking).

Importantly for our later development, LinUCB can be interpreted in Bayesian terms in the following sense: the current mean and covariance estimates $\theta \sim \gauss(\hat{\theta}_t , \Sigma_t)$ correspond to the posterior distribution obtained by combining the prior $\gauss(0, \lambda^{-1} I_{d} )$ with the likelihood $r_{t , a} \sim \gauss (\langle \phi(x_{t, a}), \theta  \rangle , 1)$.
Since 
\begin{equation}
  \langle \phi(x_{t, a}), \theta \rangle \sim \gauss \bigl(\langle \phi(x_{t, a}), \hat{\theta}_{t-1}  \rangle, \| \phi(x_{t, a}) \|_{\Sigma_{t-1}}\bigr)\, ,
\end{equation}
%
the selection rule employed by LinUCB can be viewed as selecting the action with the highest posterior $\Phi (\sqrt{\beta_t})$-th quantile, where $\Phi$ is the cumulative distribution function (CDF) of the standard normal distribution.
In other words, the maintained estimates define a \emph{confidence set} for the true parameter $\theta$, and LinUCB chooses the best action compatible with this set.

As mentioned, we will assume that each of the nominators and the ranker use the LinUCB algorithm also in the two-stage setup.\footnote{While not in line with the usual heterogeneity of the nominator algorithms, we believe this setting captures much that is salient to the interaction between general simultaneously learning ranker and nominators. It also goes one step beyond the setup in \citep{ma2020off} by considering the existence of more than one nominator.}
Since the nominators often need to rapidly sift through millions of items, we assume \emph{only} the ranker has access to the true but expensive to compute embeddings $\phi$, while each nominator $n \in [N]$ uses computationally cheaper embeddings $\phi_n(x) \in \mathbb{R}^{d_n}$.
In the rest of this document, we use the term \emph{naive} two-stage LinUCB to refer to the algorithm where each nominator \emph{independently} maintains its own estimates $\hat{\theta}_{n,t}$, $\Sigma_{n,t}$ defined as in \Cref{eq:blr_posterior_params} with $\phi$ replaced by $\phi_n$, nominate actions analogously to \Cref{eq:action_selection}, and update their posterior \emph{only} with $r_t$ and $\phi_n (x_{t, a_t})$, where we recall $a_t$ need not equal $a_{t, n}$.
Moreover, the ranker independently maintains its own estimates $\hat{\theta}$, $\Sigma_{t}$ used to select the item ultimately served to the user.

In the next section, we will first show on a simple example that such independently maintained uncertainty estimates lead to suboptimal global performance.
We then propose a solution based on synchronization of the upper bound estimates $\text{UCB}_t (a)$ used in \Cref{eq:action_selection} between the ranker and the nominators.

%% file: content/03algo.tex
\section{Coordinated exploration}
\label{sec:algo}


To understand when the \emph{naive} two-stage LinUCB implementation does not work, it is useful to know when it does. 
In particular, consider the case when all the nominators are allowed to use the \emph{same features} as the ranker $\phi_n = \phi$, and employ the \emph{same prior} $\gauss (0, \lambda_n^{-1} I_d)$ with $\lambda_n = \lambda$.
It is not hard to see that in this case $\hat{\theta}_{n, t} = \hat{\theta}_t$ and $\Sigma_{n, t} = \Sigma_{t}$ for all $t \in [T]$, and thus each nominator selects the same action as would be selected by the ranker constrained to the same action pool $A_n$.
Since $\max \{ c_1, \ldots, c_k \} = \max \{ \max \{ c_1 , \ldots , c_{k_1} \} , \ldots , \max \{ c_{k_{N - 1} + 1} , \ldots , c_k \} \}$ for any partition of $c_1, \ldots , c_k \in \R$, this then implies that the naive two-stage system behaves \emph{exactly} as a single-stage LinUCB with access to all actions would.

Because we know single-stage LinUCB is close to optimal, the above implies that any potential increase in regret must come from either the already discussed mismatch of the embeddings inherent to two-stage systems, or mismatch of the prior. 
The latter is then most often caused by the ranker being deployed for much longer time and thus better trained.
Such a scenario is common in many contemporary industrial practices \citep{covington2016deep,ma2020off}, and can be modelled in our setup by using a ranker prior with lower initial uncertainty then the nominators.
As demonstrated next, naive two-stage LinUCB is poorly equipped to handle such discrepancies.

\def\figfrac{0.24}
\begin{figure*}
  \centering
  \textbf{Noise level $\sigma_{\theta_\star} = 0.1$:} $\hat{\theta}_0 \sim \gauss (\theta_\star, 10^{-2})$\\
  \hrulespace{0.5mm}
  \vspace{1mm}
  \hspace{2cm}$\gamma = 1$ \hfill $\gamma = 10$ \hfill $\gamma = 25$ \hfill $\gamma = 50$ \hspace{1.3cm}\phantom{}
  \\
  \includegraphics[width=\figfrac\textwidth]
  {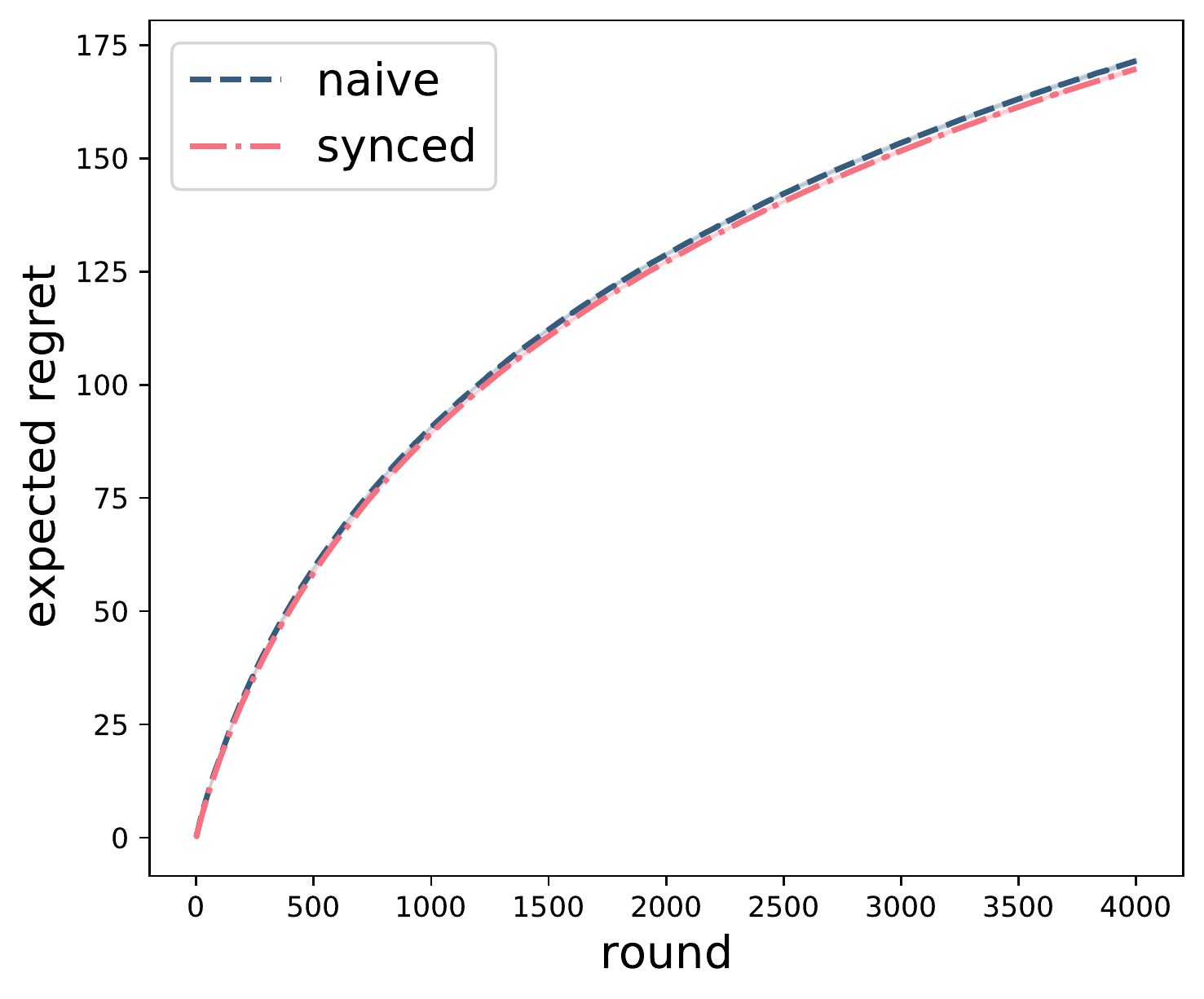}
  \hfill
  \includegraphics[width=\figfrac\textwidth]
  {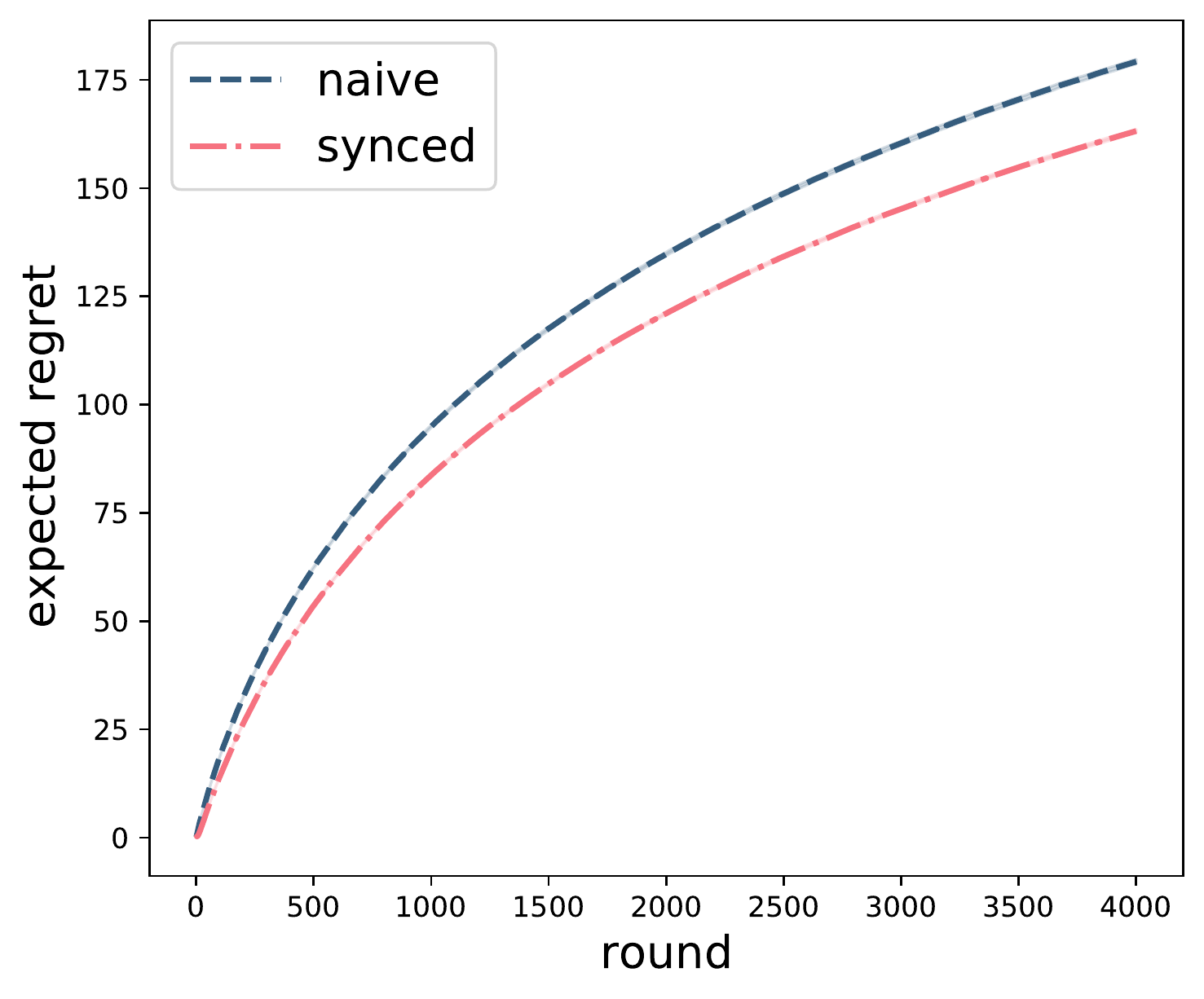}
  \hfill
  \includegraphics[width=\figfrac\textwidth]
  {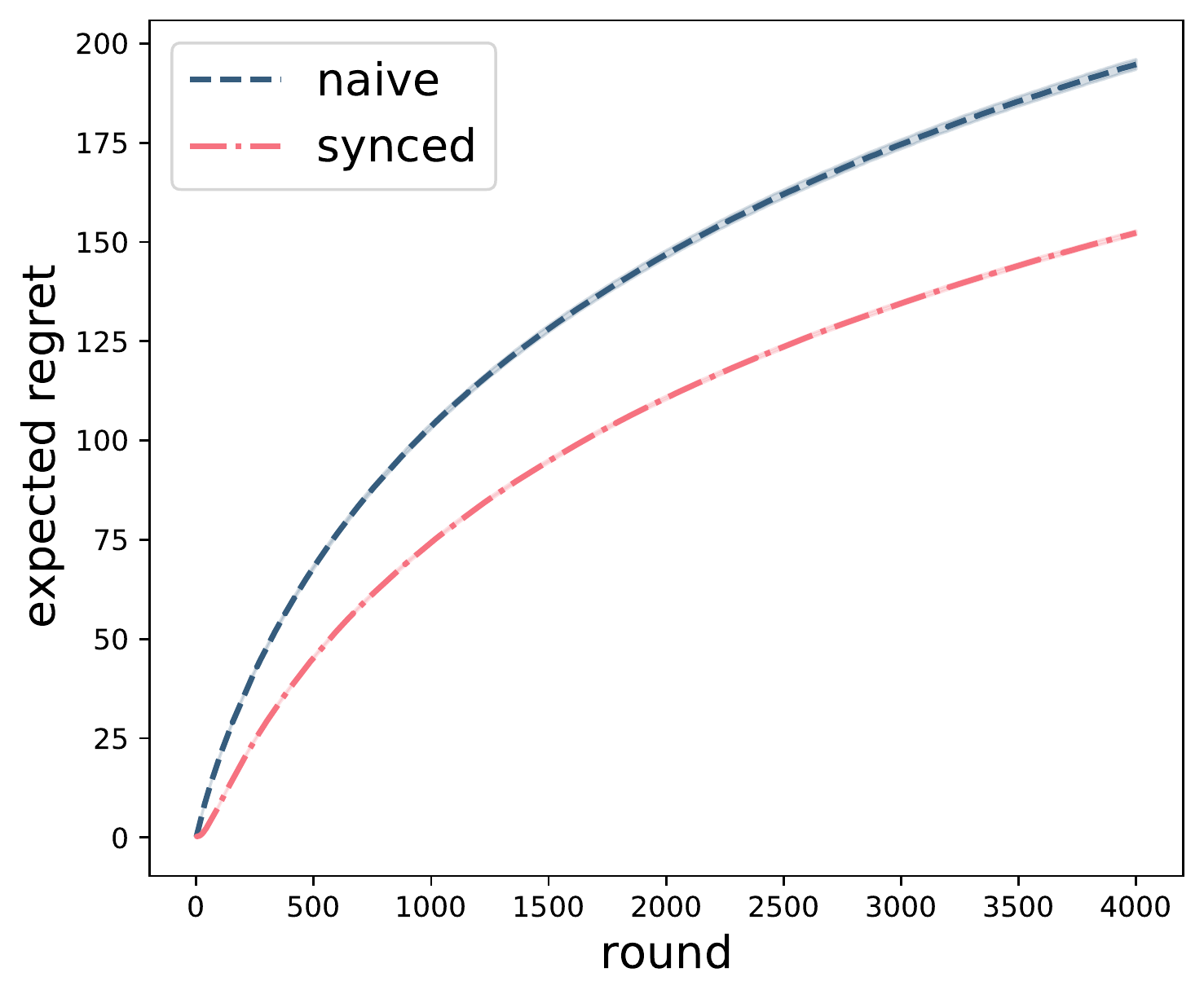}
  \hfill
  \includegraphics[width=\figfrac\textwidth]
  {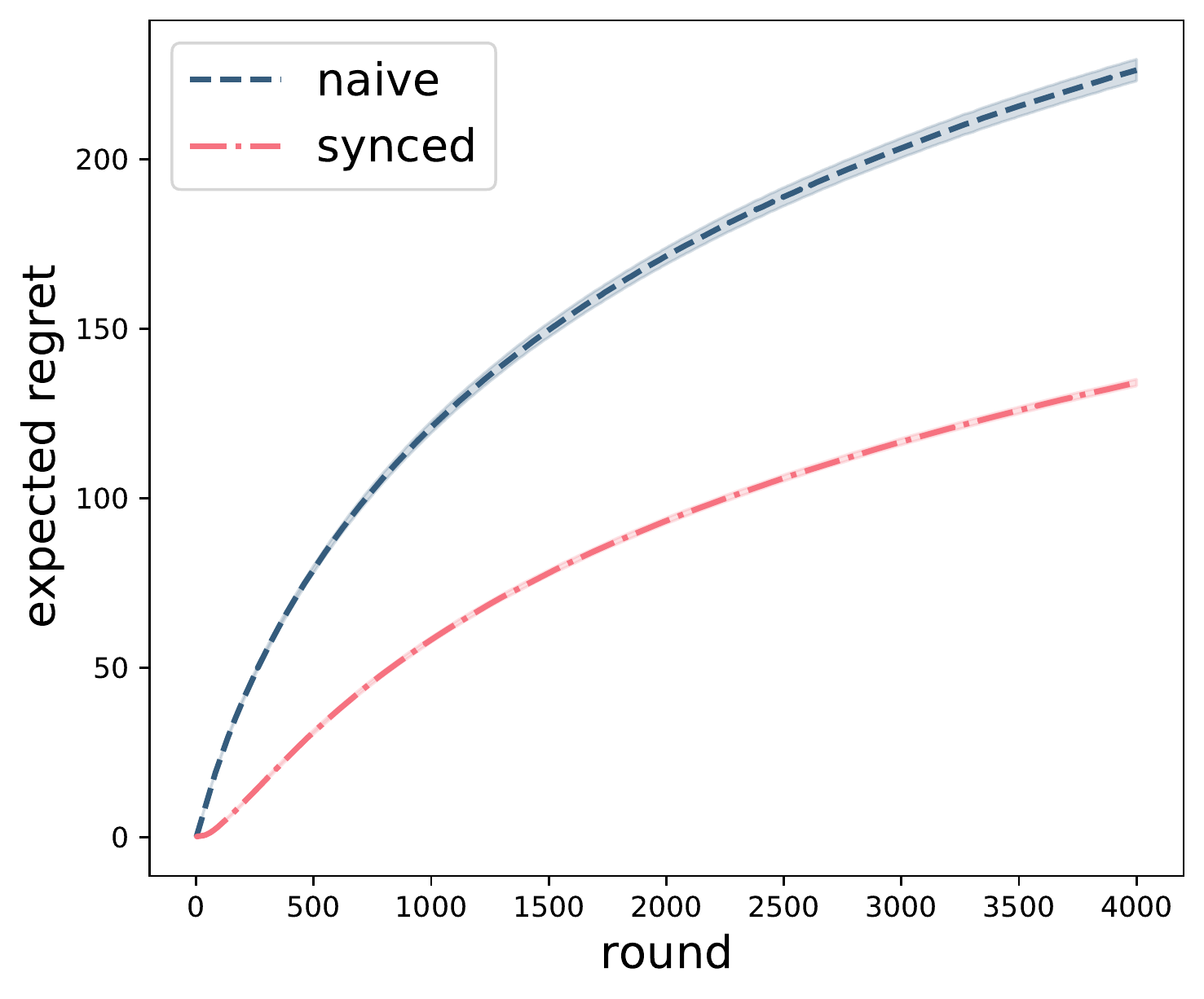}
  \\
  \textbf{Noise level $\sigma_{\theta_\star} = 0.2$:} $\hat{\theta}_0 \sim \gauss (\theta_\star, 5^{-2})$\\
  \hrulespace{0.5mm}
  \vspace{1mm}
  \hspace{2cm}$\gamma = 1$ \hfill $\gamma = 10$ \hfill $\gamma = 25$ \hfill $\gamma = 50$ \hspace{1.3cm}\phantom{}
  \\
  \includegraphics[width=\figfrac\textwidth]
  {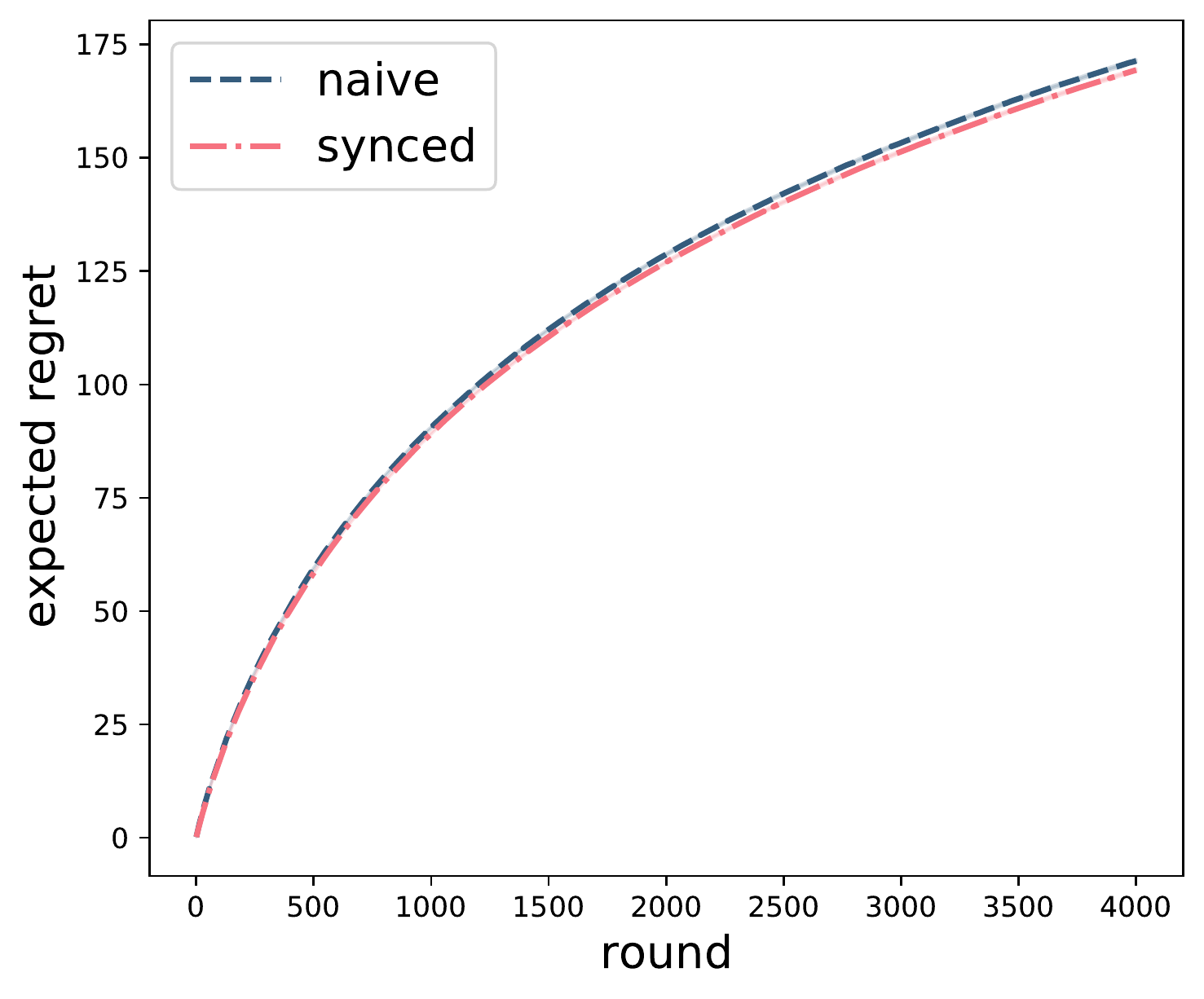}
  \hfill
  \includegraphics[width=\figfrac\textwidth]
  {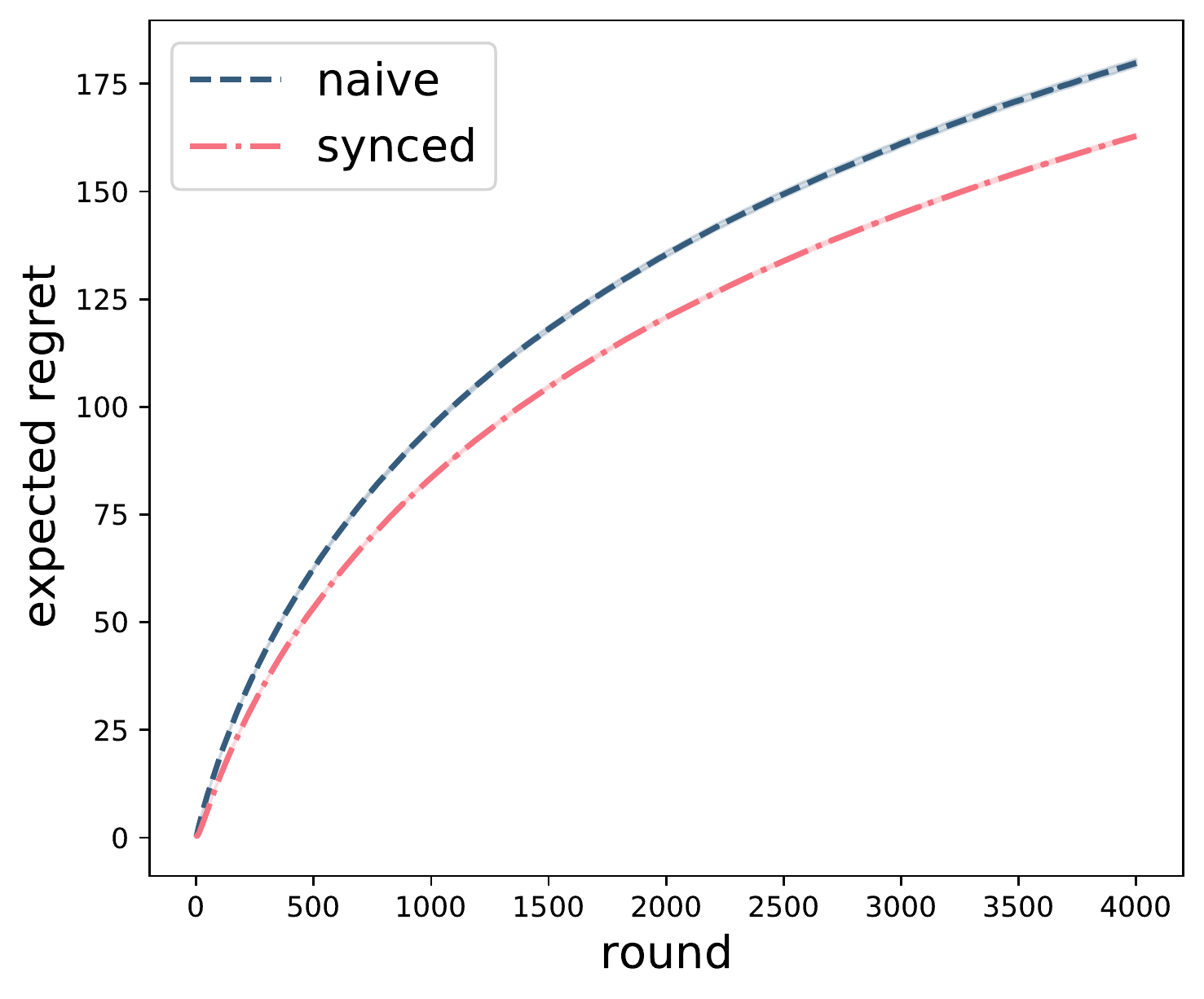}
  \hfill
  \includegraphics[width=\figfrac\textwidth]
  {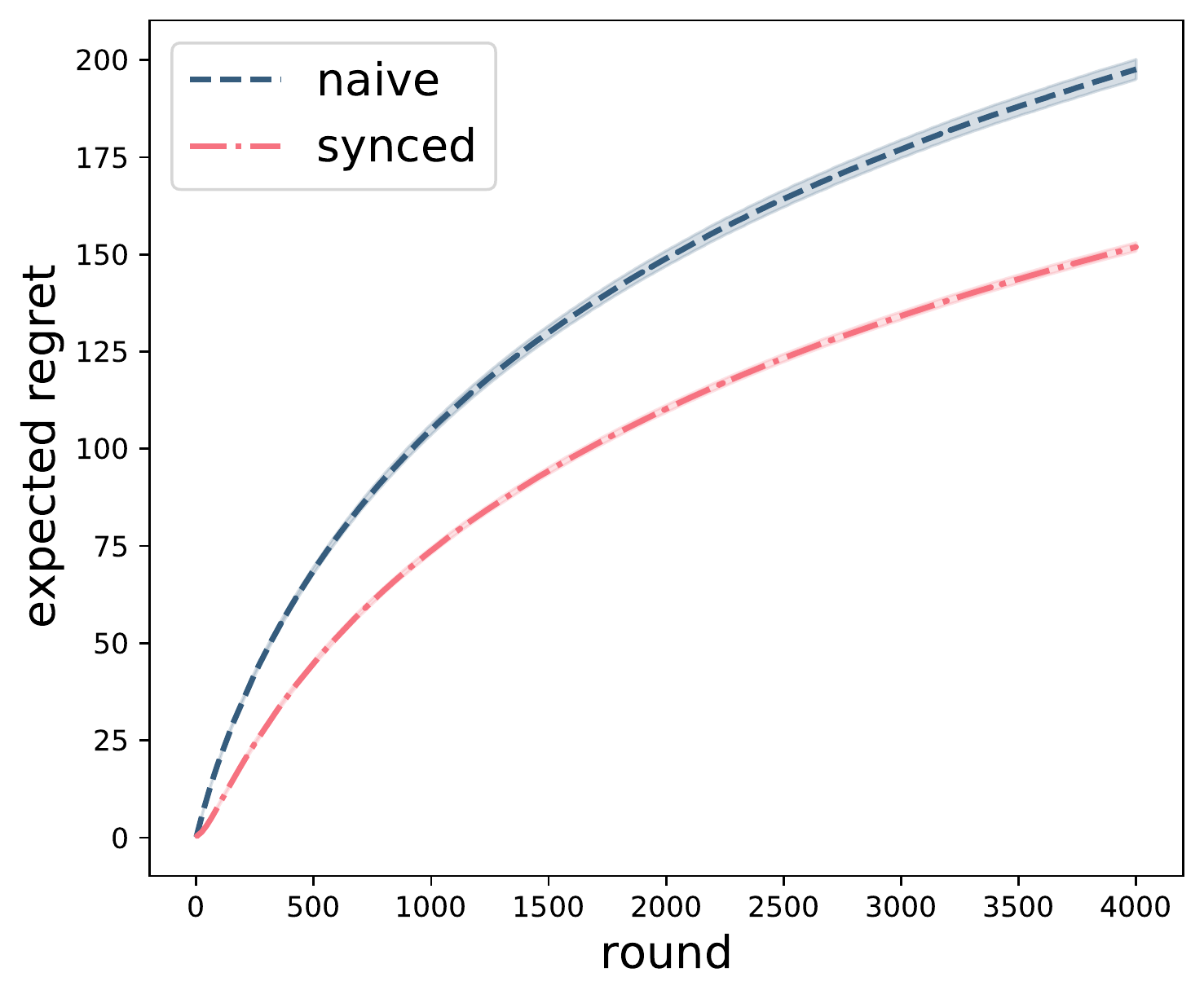}
  \hfill
  \includegraphics[width=\figfrac\textwidth]
  {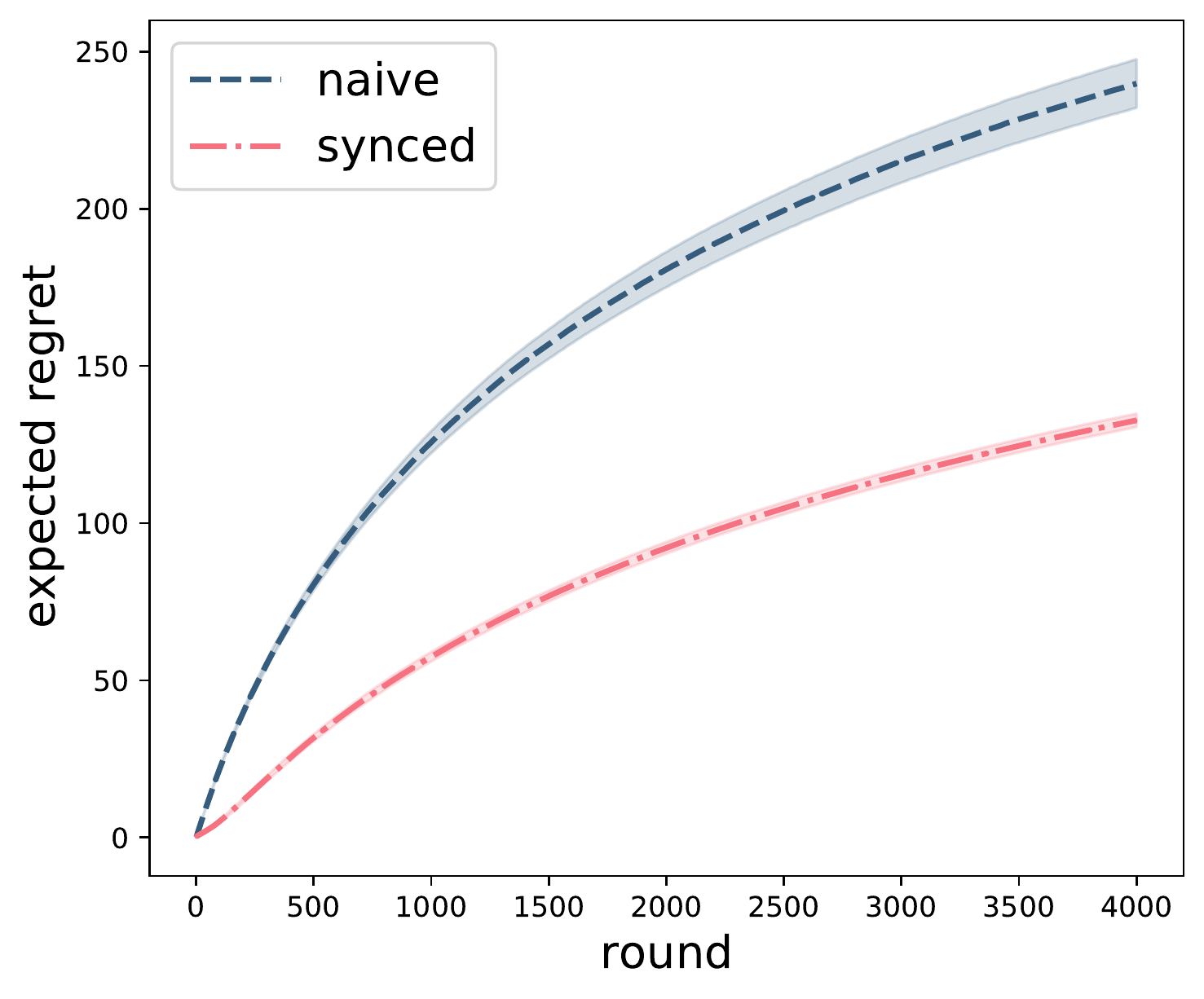}
  \caption{Naive vs.\ synchronized 2-stage recommendation. Setup described in \Cref{sect:motivating_example}. Expected regret and its 2-sigma confidence intervals were estimated over 400 runs. The level of pretraining of the ranker $\gamma$ has outsized effect on the naive but not the synchronized two-stage LinUCB, overcoming the `deadlock' effect.}\label{fig:linUCB_toy}
\end{figure*}

\begin{algorithm}[tbp]
\caption{\label{alg:sync_linUCB}Two-stage {\color{SteelBlue}synchronized} LinUCB. Here $\phi_t \coloneqq \phi(a_t)$ and $\phi_{n, t} \coloneqq \phi(a_{n, t})$ for all $n, t$.}
\footnotesize
 \textbf{Inputs:} $\hat{\theta}_0 \, , \Sigma_0 \, , (\beta_{t})_{t} \, ;\; \forall n \colon \hat{\theta}_{n , 0} \, , \Sigma_{n, 0} , (\beta_{n , t})_t$ \\
 \For{$t=1,2,\ldots, T$}{
    $\forall n \colon a_{n, t} \gets \argmax_{a \in A_n} \text{UCB}_{n, t} (a)$ \\
    $a_t \gets \argmax_{a \in \{ a_{1, t}, \ldots , a_{N, t} \}} \text{UCB}_{t} (a)$ \\
    $\Sigma_t^{-1} \gets \Sigma_{t-1}^{-1} + \phi_t \phi_t^\top$ \\
    $\hat{\theta}_t \gets \Sigma_t \left(\Sigma_{t-1}^{-1} \hat{\theta}_{t-1} + r_t \phi_t\right)$ \\
   \For{$n=1, 2, \ldots, N$} {
      $\Sigma_{n, t}^{-1} \gets \Sigma_{n, t-1}^{-1} + \phi_{n,t} \phi_{n, t}^\top$ \\
      $\hat{\theta}_{n, t} \gets \Sigma_{n, t} \left(\Sigma_{n, t-1}^{-1} \hat{\theta}_{n, t-1} + r_t \phi_{n, t}\right)$ \\
      {\color{SteelBlue}
      \uIf{$\| \phi_n (a_{n, t}) \|_{\Sigma_{n, t}} > \| \phi (a_{n, t}) \|_{\Sigma_{t}}$}{
          $\hat{\theta}_{n, t} \gets \hat{\theta}_{n, t} + \frac{\langle\hat{\theta}_{t}, \phi_{t}\rangle - \langle \hat{\theta}_{n, t}, \phi_{n, t}\rangle}{\|\phi_{n, t}\|_{\Sigma_{n, t}}^2}\Sigma_{n, t}\phi_{n, t}$ \\
          $\Sigma_{n, t}^{-1} \gets \Sigma_{n,t}^{-1} + \bigl( \frac{1}{\|\phi_t\|_{\Sigma_{t}}^{2}} - \frac{1}{\|\phi_{n, t}\|_{\Sigma_{n, t}}^{2}} \bigr)
          \phi_{n, t}\phi_{n, t}^{\top}$
      }
      }
   }
 }
\end{algorithm}

\subsection{Motivating example}\label{sect:motivating_example}


Consider a setting with \emph{only one} context, two nominators, three actions split between them as $A_1 = \{ a_1 \}$, $A_2 = \{ a_2, a_3 \}$, $\phi$ returning \emph{one-hot encodings} of the actions, and $\phi_n = \phi$ for all nominators.
The expected rewards from $a_1$ to $a_3$ are $[1/2, 1/4, 3/4] = \theta_*$ (one-hot action encodings),
and observed rewards are generated by adding i.i.d.\ Gaussian noise $\gauss (\theta_\star, 10^{-2} I_3)$.
The ranker's parameters are initialized to $\hat{\theta}_0 \sim \gauss (\theta_\star, \sigma_{\theta_\star}^2 I_3)$, and $\Sigma_{0} = (\lambda + \gamma)^{-1} I_{3}$ where $\gamma$ represents how many more samples per action the ranker has seen at $t = 1$ compared to the nominators.
For both nominators $n \in [2]$, we take $\hat{\theta} = 0$, and $\Sigma_{n, 0} = \lambda_n^{-1} I_{3}$ with regularization parameter (prior precision) $\lambda_n = \lambda = 10^{-3}$. 

To see what can go wrong in this scenario, consider the extreme case $\sigma_{\theta_\star}^2 = 0$ and $\gamma \gg 0$, i.e., the ranker has seen enough data to essentially recover the true parameter.
Since $A_1 = \{ a_1 \}$, such a ranker always picks $a_3$ (the best action) when $a_{2, t} = a_3$, and $a_1$ otherwise.
In the naive implementation, this results in a \emph{`deadlock'} where the
second nominator's uncertainty about $a_2$ never decreases, leading it to mostly nominate $a_2$ over $a_3$, entailing \emph{linear} regret.
While the extreme case may be rare in practice, \Cref{fig:linUCB_toy} shows the effect remains significant even when the ranker is not fully trained, i.e., $\sigma_{\theta_\star} > 0$, and $\gamma > 1$ but not overly large.



\subsection{Synchronized two-stage LinUCB}

The deadlock observed in the previous section is due to a lack of communication of uncertainty between the ranker and the nominators. 
Since the computational constraints inherent to the two-stage setup entail distinct embedding functions, the issue cannot be addressed by simply setting nominator parameters to those of the ranker.
However, because action selection only depends on the estimated marginal quantiles of the \emph{rewards}, we can update the nominator's estimates based on the reward statistics computed by the ranker.

We propose to \emph{synchronize} each nominator $n$ in the rounds where $\| \phi_n (a_{n, t}) \|_{\Sigma_{n, t}} > \| \phi (a_{n, t}) \|_{\Sigma_{t}}$, i.e., when the nominator is more uncertain about its selected action than the ranker (the cause of the `deadlock' in \Cref{sect:motivating_example}).
In particular, we want to minimally adjust the nominator posterior so that it matches the ranker's mean and variance which fully determine $\text{UCB}_{n, t}(a_{n, t})$ (see \Cref{sec:setup}).
Defining the minimality in terms of Kullback-Leibler (KL) divergence, this can be achieved by solving the constrained optimization problem:
\begin{align}\label{eq:kl_sync}
    \min_{m, S}
    \quad
    &
        \text{KL}\left(\mathcal{N}\left(m, S\right)\;\middle \|\;\mathcal{N}\left(\hat{\theta}_{n, {t}}, \Sigma_{n, {t}}\right)\right)
    \\
    \text{subject to}
    \quad
    &\langle m, \phi_{n} (a_{n, t}) \rangle = \langle \hat{\theta}_{t}, \phi (a_{n, t}) \rangle
    \, , \qquad\,\,
    {\color{gray}\text{\small(mean)}}
    \nonumber
    \\
    &\|\phi_n (a_{n, t})\|_{S} = \|\phi (a_{n, t})\|_{\Sigma_{t}}
    \, , \quad
    {\color{gray}\text{\small(covariance)}}
    \nonumber
\end{align}
When synchronization is performed after the usual update, the solution to \Cref{eq:kl_sync} gives us \Cref{alg:sync_linUCB}.

\begin{figure}
  \centering
  \includegraphics[width=0.8\columnwidth]{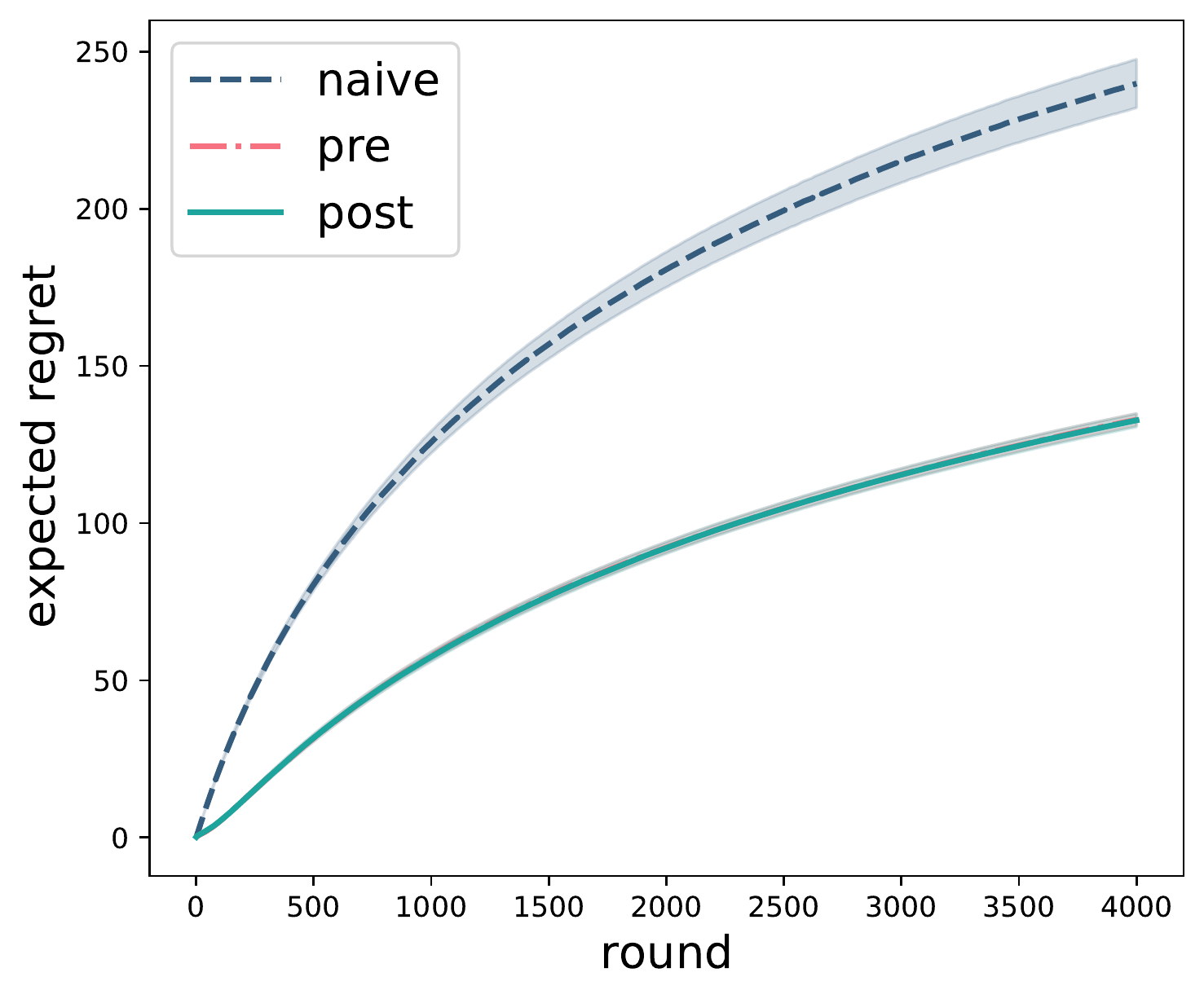}
  \caption{Comparison of post- vs.\ pre-update synchronization for a noise level of $0.2$ and $\gamma = 50$.
  In this setting, the line for pre-update synchronization is barely visible, because it is covered by the line for post-update synchronization.}
  \label{fig:post_vs_pre}
\end{figure}

Note that the selected KL divergence penalizes overdispersion compared to the previous distribution, meaning the resulting replacement for $\gauss(\hat{\theta}_{n, {t}}, \Sigma_{n, {t}} )$ should not have more uncertainty. 
Furthermore, if we modify \Cref{alg:sync_linUCB} to perform the synchronization before the usual update---i.e., swap the black and blue lines within the inner-most for-loop and replace the if-condition by $\| \phi_n (a_{n, t}) \|_{\Sigma_{n, t-1}} > \| \phi (a_{n, t}) \|_{\Sigma_{t-1}}$---we arrive at an algorithm that only uses quantities already computed during selection of $a_{n, t}$ and $a_t$, minimizing the additional computation required; 
in experiments, both versions of the algorithm performed essentially the same, which we show in an example setting in \Cref{fig:post_vs_pre}.
\nk{For a future version we should probably only show the compared lines (not naive) and make them somewhat transparent. Or perhaps even show absolute difference on log scale?}

Finally, let us consider our synchronized two-stage LinUCB algorithm in the context of the motivating example from \Cref{sect:motivating_example}.
Because we assumed $\phi = \phi_n$ are one-hot encodings of the actions, we see that the ranker variance for an action $j \in [3]$ at time $t$ is $\| \phi(a_j) \|_{\Sigma_t}^2 = (\lambda + \gamma + n_{tj})^{-1}$ where $n_{tj}$ is the number of times the ranker selected $a_j$.
The \emph{first} time a nominator select $a_j$, its variance before the round update will be $\| \phi_n(a_j) \|_{\Sigma_{n, t}}^2 = (\lambda + n_{tj})^{-1}$.\footnote{Since we allow the pools $A_n$ to be overlapping, $n_{tj}$ could generally be greater than zero here. This is not the case for the example from \Cref{sect:motivating_example} though.}
Inspecting the synchronization update for $\Sigma_{n, t}^{-1}$ in blue (\Cref{alg:sync_linUCB}), the new value for the $(jj)$\textsuperscript{th} entry of $\Sigma_{n, t}^{-1}$ amounts to
\begin{align*}
    \underbrace{\lambda + n_{tj}}_{\Sigma_{n, t}^{-1}}
    + 
    \underbrace{\lambda + \gamma + n_{tj}}_{
      \| \phi(a_j) \|_{\Sigma_t}^{-2}
    }
    -
    \underbrace{(\lambda + n_{tj})}_{
      \| \phi_n(a_j) \|_{\Sigma_{n, t}}^{-2}
    }
    = 
    \| \phi(a_j) \|_{\Sigma_t}^2
    \, ,
\end{align*}
while the other variances will remain as in $\Sigma_{n, t}$ at time $t$.
Since an analogous claim holds for the mean $\hat{\theta}_{n, t}$, we conclude the synchronization update ensures the posteriors of the ranker and the nominators match after each nominator \emph{selected} (but not necessarily seen \emph{recommended}) each $a \in A_n$ exactly once.
Because the posteriors never diverge after they are fully matched, our algorithm starts behaving like single-stage LinUCB from thereon, which we know is near optimal for the task.
This is confirmed in \Cref{fig:linUCB_toy} where the more pretrained the ranker is (higher $\gamma$), the better the synchronized and the worse the naive LinUCB do.


%% file: content/05conclusion.tex
\section{Conclusion}
\label{sec:conclusion}

We have shown that naive deployment of LinUCB in two-stage recommender systems may result in suboptimal performance, and close to linear regret in extreme cases.
The suboptimality is due to the mismatch between both the embeddings used by the ranker and the nominators, and the gap between the amount of training data the ranker has seen compared to the nominators.
Both of these issues are inherent to the setting in which two-stage recommenders are typically deployed in industry, and thus pose important barriers to achieving a good exploration-exploitation trade-off. 

We have proposed a simple modification of the LinUCB algorithm based on communication of inferred statistics between the ranker and the nominators. 
Our algorithm can be implemented with minimal computational overhead, and achieves superior empirical results compared to the naive two-stage LinUCB implementation.
While focusing solely on LinUCB, we suspect the `deadlock' problem identified in \Cref{sect:motivating_example} is pertinent to any exploration algorithm which in part selects its actions based on the level of uncertainty about them (e.g., all `optimism in the face of uncertainty' type algorithms, \citealp{lattimore2020bandit}).
Since the principles motivating \Cref{eq:kl_sync} do not hinge on LinUCB in particular, we hope these issues could also be addressed by communication of statistics from the ranker.

This preprint is based on a workshop  version of our work. 
We plan to publish an extended version with additional experiments on real-world data and extended discussion in the near future.